\documentclass[aps,showpacs,preprint,graphics]{revtex4}
\usepackage{amssymb}
\usepackage[dvips]{graphicx}
\usepackage{amsmath}
\usepackage{bm}
\usepackage{epsfig}

\begin{document}

\title{Gravitational waves from a Weyl-Integrable manifold: a new formalism.}

\author{ $^{1}$ Jes\'us Mart\'in Romero\thanks{E-mail address: jesusromero@conicet.gov.ar},
$^{1,2}$ Mauricio Bellini \thanks{E-mail address: mbellini@mdp.edu.ar} and $^{3}$ Jos\'e Edgar Madriz Aguilar
\thanks{E-mail address: madriz@mdp.edu.ar} }
\affiliation{ $^{1}$ Departamento de F\'isica, Facultad de Ciencias Exactas y Naturales\\
    Universidad Nacional de Mar del Plata, Funes 3350, C.P. 7600, Mar del Plata, Argentina\\
     $^{2}$ Instituto de Investigaciones F\'isicas de Mar del Plata (IFIMAR)\\
     Consejo Nacional de Investigaciones Cient\'ificas y T\'ecnicas (CONICET), Argentina.\\
     E-mail: jesusromero@conicet.gov.ar, mbellini@mdp.edu.ar\\
     $^{3}$ Departamento de Matem\'aticas, Centro Universitario de Ciencias Exactas e ingenier\'{i}as (CUCEI),
Universidad de Guadalajara (UdG), Av. Revoluci\'on 1500 S.R. 44430, Guadalajara, Jalisco, M\'exico.  \\
E-mail: madriz@mdp.edu.ar, edgar.madriz@red.cucei.udg.mx}

\begin{abstract}
We study the variational principle over an
Hilbert-Einstein like action for an extended geometry taking into account torsion
and non-metricity. By extending the semi-Riemannian geometry, we obtain an
effective energy-momentum tensor which can be interpreted as physical sources.
As an application we develop a new
manner to obtain the gravitational wave equations on a Weyl-integrable
manifold taking into account the non-metricity and non-trivial
boundary conditions on the minimization of the action, which can
be identified as possible sources for the cosmological constant
and provides two different equations for gravitational waves. We examine gravitational waves
in a pre-inflationary cosmological model.
\end{abstract}

\maketitle

\vskip .5cm
Torsion, nonmetricity, Weyl-Integrable gravity, gravitational waves, inflation.

\section{Introduction}

In the standard treatment to minimize the action, when a manifold
has a boundary $\partial{\cal{M}}$, the action should be
supplemented by a boundary term, so that the variational principle
to be well-defined\cite{Y,GH}. However, this is not the only
manner to study this problem. As was recently
demonstrated\cite{06}, there is another way to include the flux
around a hypersurface that encloses a physical source without the
inclusion of another term in the Hilbert-Einstein (HE) action.
This treatment imposes a constraint on the dynamics obtained by
varying the EH action. In that paper was demonstrated that the
non-zero flux of the vector metric fluctuations through the closed
3D Gaussian-like hypersurface, is responsible for the
gauge-invariance of gravitational waves. In present paper
we are dealing  with the variational principle over a Hilbert-Einstein like
action, using an extended geometry with torsion and non-metricity, from which
we obtain an effective energy-momentum tensor with sources in the
torsion and the non-metricity. It can be viewed in a Riemannian geometry as a describing an effective stress tensor
that represents a geometrically
induced matter. Additionally,  we develop a
new manner to obtain gravitational waves on a Weyl-integrable
manifold, which has non-metricity and nontrivial boundary terms
included. 

The paper is organised as follows: in the following section we shall study the general formalism. In Sect. IV we examine the formalism in absence of torsion, taking into account
purely Weylian contributions. In Sect. V we deal only with the contributions due to boundary terms.
In Sect. VI we examine an example in which
massless gravitons are emitted during a pre-inflationary epoch of the universe. Finally, in Sect. VII we develop some
final remarks.

\section{General formalism}

We consider the variational
principle in presence of torsion and non-metricity in an Hilbert-Einstein
action. We shall start by considering an action in an extended geometry
(i.e. a non-Riemannian manifold) conformed by a
gra\-vi\-ta\-tio\-nal sector without the presence of matter in
such generalized geometry. The Hilbert-Einstein action was
extensively studied in Riemannian geometry \cite{01}, but we shall
deal with an extended geometry:
\begin{equation}\label{a1}
{\cal S}=\frac{1}{2\kappa} \int_{V}d^{4}x\sqrt{-g}\,{\cal R},
\end{equation}
where $V$ denotes the volume of a spacetime manifold featured by a
non-metricity \cite{01.1},\cite{02} and a general torsion
\cite{01.1},\cite{03}. Furthermore, $g$ is the determinant of the
metric tensor $g_{\alpha\beta}$, the gravitational coupling is
denoted by $\kappa=8\pi G$ and ${\cal R}=g^{\alpha\beta}R_{\alpha\beta}$
is the scalar curvature.  For a coordinate basis of the tangent
space $\lbrace
\partial_{\sigma}\rbrace$, the components of the Riemann tensor
are given by
\begin{equation}\label{a2}
R^{\alpha}_{\beta\mu\nu}=\Gamma^{\alpha}_{\beta\nu ,\mu}-\Gamma^{\alpha}_{\beta\mu ,\nu}
+\Gamma^{\sigma}_{\beta\nu}\Gamma^{\alpha}_{\sigma\mu}-\Gamma^{\sigma}_{\beta\mu}\Gamma^{\alpha}_{\sigma\nu},
\end{equation}
where the symbols $\Gamma^{\alpha}_{\mu\nu}$ denote the coordinate
components for a generalized connection defined by
\cite{01.1},\cite{04}
\begin{equation}\label{a3}
\nabla_{\partial_\alpha}\partial_{\beta}=\Gamma^{\epsilon}_{\beta\alpha}\partial_{\epsilon}.
\end{equation}
These components can be written in the general form
\begin{equation}\label{a4}
\Gamma^{\sigma}_{\mu\nu}=\left\lbrace \,^{\sigma}_{\mu\nu}\right\rbrace+K^{\sigma}_{\mu\nu},
\end{equation}
where $\left\lbrace \,^{\gamma}_{\alpha\beta}\right\rbrace$ are
the components of the usual Riemannian connection (the second kind
Christoffel symbols) and $K^{\sigma}_{\mu\nu}$ is a contortion
tensor due to torsion and non-metricity, defined by \cite{05}
\begin{small}
\begin{equation}\label{a5}
K^{\sigma}_{\mu\nu}=-\frac{g^{\beta\sigma}}{2}\{\tau^{\alpha}_{\nu\beta}\,g_{\mu\alpha}
+\tau^{\alpha}_{\mu\beta}\,g_{\alpha\nu}-\tau^{\alpha}_{\nu\mu}\,g_{\alpha\beta}+N_{\mu\beta\nu}+N_{\beta\nu\mu}-N_{\mu\nu\beta}\},
\end{equation}
\end{small}
such that $\tau^{\alpha}_{\mu\nu}$ and $N_{\alpha\beta\gamma}$ are respectively the torsion and
the non-metricity tensors. For  a
coordinate basis, they are
\begin{eqnarray}\label{a6}
\tau^{\alpha}_{\mu\nu}&=&\Gamma^{\alpha}_{\mu\nu}-\Gamma^{\alpha}_{\nu\mu},\\
\label{a7}
N_{\alpha\beta\gamma}&=& g_{\beta\gamma ;\alpha},
\end{eqnarray}
where the semicolon denotes the covariant derivative defined in terms of
the $\Gamma$ connection (i.e. defined on the extended manifold).

Now, in order to derive the dynamical equations for gravitational
waves on this general space-time manifold in a novel and
consistent manner, we shall use a variational procedure.
The variation of the action (\ref{a1}) leaves to the expression
\begin{equation}\label{a10}
\delta\left[\sqrt{-g}\,{\cal R}\right]=\sqrt{-g}\left[\delta g^{\alpha\beta} G_{\alpha\beta}+g^{\alpha\beta}\delta R_{\alpha\beta}\right],
\end{equation}
where $G_{\alpha\beta}=R_{\alpha\beta}-(1/2){\cal R} \,g_{\alpha\beta}$
is a generalization of the Einstein tensor, due to the fact that
it is calculated in terms of the $\Gamma$ connection (\ref{a4}).
Therefore, it is easy to show that the generalized Einstein tensor
contains contributions associated with both tensors, torsion and non-metricity.
The last term between brackets of the equation (\ref{a10}) can be
written as a generalized Palatini's identity in the form
\begin{equation}\label{a11}
 g^{\alpha\beta}\,\delta R_{\alpha\beta}={W^{\mu}}_{;\mu}-{g^{\alpha\beta}}_{;\mu}\,W^{\mu}_{\alpha\beta}
-\frac{1}{2}g^{\alpha\beta}(\delta\Gamma^{\mu}_{\sigma\beta}\,\tau^{\sigma}_{\alpha\mu}+\delta\Gamma^{\mu}_{\sigma\alpha}\,\tau^{\sigma}_{\beta\mu}),
\end{equation}
where $W^{\mu}=g^{\alpha\beta}\,W^{\mu}_{\alpha\beta}$. Here, we
have introduced the auxiliary tensor $W^{\mu}_{\alpha\beta}$
defined by
\begin{equation} \label{a12}
W^{\mu}_{\alpha\beta}=\,\delta\Gamma^{\mu}_{\alpha\beta}-\,\delta\Gamma^{\sigma}_{\sigma\beta}\,\delta^{\mu}_{\alpha}.
\end{equation}
Inserting (\ref{a11}) in (\ref{a10}), and using the identity
$g^{\alpha\nu}_{;\mu}=-g^{\beta\nu}\,g^{\alpha\sigma}\,N_{\sigma\beta\mu}$,
we obtain the variation of the gravitational sector of the action
(\ref{a1})
\begin{equation}
\delta {\cal S}=\int_{V} d^{4}x \sqrt{-g} \, G_{\alpha\beta}\delta
g^{\alpha\beta} +\int_{ V}d^{4}x \sqrt{-g}\, W^{\mu}_{;\mu}
+\int_{V}d^{4}x \sqrt{-g}\,
N_{\mu\alpha\beta}W^{\beta\mu\alpha}-\frac{1}{2}\int_{V}d^{4}x
\sqrt{-g}\,\zeta_{\alpha\beta}g^{\alpha\beta}, \label{a13}
\end{equation}
where $\zeta_{\alpha\beta}$ is an auxiliary tensor field, given by
\begin{equation}\label{aa13}
\zeta_{\alpha\beta}=\delta\Gamma^{\mu}_{\sigma\beta}\tau^{\sigma}_{\alpha\mu}+\delta\Gamma^{\mu}_{\sigma\alpha}\tau^{\sigma}_{\beta\mu}.
\end{equation}
The third and fourth integrals in (\ref{a13}), are respectively related
to the presence of non-metricity and torsion. On the
other hand, the second integral in (\ref{a13}) can be reduced to a
3D hypersurface integral, in virtue of the Stokes Theorem:
\begin{eqnarray}\label{super}
\int_{ V}d^{4}x \sqrt{-g}\,
W^{\mu}_{;\mu}=\int_{\partial V}d^{3}x \sqrt{-g}\,
W^{\mu}n_{\mu},
\end{eqnarray}
where $n_{\mu}$ is a vector field which is normal to the
hypersurface $\partial V$. It is usual in the literature to
suppose that the surface integral must be neglected when the
radius of $\partial V$ is large enough to impose that the field
$W^{\mu}\rightarrow 0$ in such limit, or when $W^{\mu}$ is tangent
to $\partial V$, that is, when $W^{\mu}$ satisfies the relation
$W^{\mu}n_{\mu}=0$. In this paper we shall adopt a different path,
and we shall see that the 3D hypersurface term is a source for the
cosmological constant \cite{06}. From the expression
(\ref{super}), can be noticed that the term
$W^{\mu}_{\,;\,\nu}=g^{\alpha\beta}_{\,;\,\nu}\,W^{\mu}_{\alpha\beta}+g^{\alpha\beta}\,W^{\mu}_{\alpha\beta\,;\,\nu}\neq
g^{\alpha\beta}\,W^{\mu}_{\alpha\beta\,;\,\nu}$ has contributions of non-metricity. 
This implies that, if we drop this term, the new
contribution is not very important. However, when we neglect such
boundary term, we must be careful with non-metricity.

We must notice that the Einstein tensor in (\ref{a13}) can be
written as a Riemannian part, plus a non-Riemannian one, in the
form
\begin{small}
\begin{equation}\label{a14}
G_{\alpha\beta}=\,\bar{G}_{\alpha\beta}+K^{\mu}_{\mu\beta\,|\alpha}-K^{\mu}_{\alpha\beta\,|\mu}+K^{\nu}_{\mu\beta}
K^{\mu}_{\nu\alpha}-K^{\nu}_{\alpha\beta}K^{\mu}_{\nu\mu}-
-\frac{g^{\sigma\gamma}}{2}(K^{\mu}_{\mu\gamma\,|\sigma}-K^{\mu}_{\sigma\gamma\,|\mu}+K^{\nu}_{\mu\gamma}
K^{\mu}_{\nu\sigma}-K^{\nu}_{\sigma\gamma}K^{\mu}_{\nu\mu})\,g_{\alpha\beta}.
\end{equation}
\end{small}
In this expression, the {\em bar} in $\bar{G}_{\alpha\beta}$
indicates that the Einstein tensor is calculated with the
Levi-Civita connections. The symbol "$|$" denotes the Riemannian
covariant derivative. It follows from (\ref{a14}) that when the
non-metricity and the torsion vanish. The Einstein tensor in the
first integral of (\ref{a13}), simply reduces to the usual
Einstein tensor calculated with the Levi-Civita connections.

Now, using the equations (\ref{a4}) and (\ref{a5}), the auxiliary
tensor $W^{\sigma}_{\alpha\beta}$ and the vector field
$W^{\sigma}$, can be written  as
\begin{small}
\begin{eqnarray}
W^{\mu}_{\alpha\beta}&=&\left[\frac{g^{\lambda\mu}}{2}\{\delta
g_{\alpha\lambda\,,\beta}+\delta g_{\lambda\beta\,,\alpha} -\delta
g_{\beta\alpha\,,\lambda}-\tau^{\rho}_{\beta\lambda}\,\delta
g_{\alpha\rho}-\tau^{\rho}_{\alpha\lambda}\,\delta
g_{\rho\beta}\}\right.\nonumber \\
&-&\frac{\delta g^{\lambda\mu}}{2}\{g_{\alpha\lambda\,,\beta}+
g_{\lambda\beta\,,\alpha}-
g_{\beta\mu\,,\lambda}-\tau^{\rho}_{\beta\lambda}\,
g_{\alpha\rho}-\tau^{\rho}_{\alpha\lambda}\,g_{\rho\beta}-N_{\alpha\lambda\beta}-
N_{\lambda\beta\alpha}+ N_{\beta\alpha\lambda}\}\nonumber
\\
&-&\left.\frac{g^{\lambda\sigma}}{4}(\delta
g_{\lambda\sigma\,,\beta}\,\delta^{\mu}_{\alpha}+\delta
g_{\lambda\sigma\,,\alpha}\,\delta^\mu_{\beta})+\frac{\delta
g^{\lambda\sigma}}{4}(g_{\lambda\sigma\,,\beta}\,\delta^\mu_\alpha+g_{\lambda\sigma\,,\alpha}\,\delta^\mu_\beta
-N_{\lambda\sigma\beta}\,\delta^\mu_\alpha-N_{\lambda\sigma\alpha}\,\delta^\mu_\beta)\right]\nonumber
\\
&+&
\left[\frac{g^{\lambda\mu}}{2}\tau^{\rho}_{\beta\alpha}\,\delta
g_{\rho\lambda}-\frac{g^{\lambda\mu}}{2}\tau^{\rho}_{\beta\alpha}\,g_{\rho\lambda}-\frac{g^{\lambda\sigma}}{4}(\delta
g_{\lambda\sigma\,,\beta}\,\delta^\mu_\alpha-\delta
g_{\lambda\sigma\,,\alpha}\,\delta^\mu_\beta)\right.\nonumber
\\
&+&\left.\frac{\delta
g^{\lambda\sigma}}{4}(g_{\lambda\sigma\,,\beta}\,\delta^\mu_\alpha-
g_{\lambda\sigma\,,\alpha}\,\delta^\mu_\beta+N_{\lambda\sigma\beta}\,\delta^\mu_\alpha-N_{\lambda\sigma\alpha}\,\delta^\mu_\beta)\right].
\label{ws} \\
W^{\mu}&=&
g^{\alpha\beta}\left[\frac{g^{\lambda\mu}}{2}\left\{\delta
g_{\alpha\lambda\,,\beta}+\delta g_{\lambda\beta\,,\alpha}-\delta
g_{\beta\alpha\,,\lambda}-\tau^{\rho}_{\beta\lambda}\,\delta
g_{\alpha\rho}-\tau^{\rho}_{\alpha\lambda}\,\delta
g_{\rho\beta}\right\}\right.\nonumber\\ &&-\frac{\delta
g^{\lambda\mu}}{2}\left\{g_{\alpha\lambda\,,\beta}+
g_{\lambda\beta\,,\alpha}-
g_{\beta\alpha\,,\lambda}-\tau^{\rho}_{\beta\lambda}\,
g_{\alpha\rho}-\tau^{\rho}_{\alpha\lambda}\,g_{\rho\beta}-N_{\alpha\lambda\beta}-
N_{\lambda\beta\alpha}\right.\nonumber\\ && \left.+
N_{\beta\alpha\lambda}\right\}-\frac{g^{\lambda\sigma}}{4}\left(\delta
g_{\lambda\sigma\,,\beta}\,\delta^{\mu}_{\alpha}+\delta
g_{\lambda\sigma\,,\alpha}\,\delta^{\mu}_{\beta}\right)+\frac{\delta
g^{\lambda\sigma}}{4}\left(g_{\lambda\sigma\,,\beta}\,\delta^{\mu}_{\alpha}
+g_{\lambda\sigma\,,\alpha}\,\delta^{\mu}_{\beta}\right.\nonumber\\
&&-\left.\left.N_{\lambda\sigma\beta}\,\delta^{\mu}_{\alpha}-N_{\lambda\sigma\alpha}\,\delta^{\mu}_{\beta}\right)\right].
\label{wn}
\end{eqnarray}
\end{small}
Notice that the term in the first bracket of the expression
(\ref{ws}) is the symmetric part of $W^{\mu}_{\alpha\beta}$, while
that the term in the second bracket is the antisymmetric part of
$W^{\mu}_{\alpha\beta}$, which do not contribute in $W^{\mu}$.

\section{The torsionless case}

The torsion contribution was historically linked to spin matter\cite{kleinert1,kleinert0}, 
or magnetic monopoles\cite{monos}. This is a controversial topic that deserves a rigorous treatment, which goes beyond the
scope of this work and we shall study in a further work. For this reason, in this
section we shall study the case of a space-time manifold equipped
with non-metricity and without torsion. Under these
considerations, the equation (\ref{a13}) reduces to
\begin{equation}\label{tn1}
\delta S=\int_{V}d^{4}x \sqrt{-g}\,\left( G_{\alpha\beta}\delta
g^{\alpha\beta}+ W^{\mu}_{;\,\mu}+
N_{\mu\alpha\beta}W^{\beta\mu\alpha}\right),
\end{equation}
where we are taking into account a boundary condition in which the
3D hypersurface integral term in (\ref{a13}) is nonzero. In the
absence of matter sources, we can associate the geometrical vacuum
with a physical one, through the introduction of a dynamical
cosmological constant term, in the form
\begin{equation}\label{tn2}
W^{\mu}_{;\mu}+N_{\alpha\beta\mu}W^{\mu\alpha\beta}=\Lambda
(x)g_{\alpha\beta}\delta g^{\alpha\beta}.
\end{equation}
Here, $\Lambda (x)$ is a function of the proper time on the
manifold defined by the connections (\ref{a4}), and can play the
role of a dynamical cosmological constant with two sources: the
surface-associated term and the non-metricity term of (\ref{tn1}).
We must see two separated contributions to the cosmological
constant
\begin{equation}\label{tn21}
N_{\alpha\beta\mu}W^{\mu\alpha\beta}=\Lambda_1
(x)g_{\alpha\beta}\delta g^{\alpha\beta},
\end{equation}\begin{equation}\label{tn22}
W^{\mu}_{;\mu}=\Lambda_2 (x)g_{\alpha\beta}\delta g^{\alpha\beta},
\end{equation} with $\Lambda(x)=\Lambda_1(x)+\Lambda_2(x)$. Hence the equation (\ref{tn1})
becomes
\begin{equation}\label{tn3}
\delta S=\int_{V}d^{4}x \sqrt{-g}\,\left(
G_{\alpha\beta}+\Lambda(x)\, g_{\alpha\beta}\right)\delta
g^{\alpha\beta}.
\end{equation}
By imposing the condition that the action is an extreme: $\delta S
=0$, we obtain
\begin{equation}\label{tn4}
R_{\alpha\beta}-\frac{1}{2}R g_{\alpha\beta}+\Lambda g_{\alpha\beta}=0,
\end{equation}
which correspond to the extended Einstein field equations in
vacuum where the Ricci tensor $R_{\alpha\beta}$ and the scalar
curvature $R$, are both calculated with the affine connections
(\ref{a4}), without torsion. Using (\ref{a14}), the equations
(\ref{tn4}) can be written as
\begin{equation}\label{tn5}
\bar{R}_{\alpha\beta}-\frac{1}{2}\,\bar{R}\,g_{\alpha\beta}+\Lambda
g_{\alpha\beta}=k\,\bar{T}_{\alpha\beta},
\end{equation}
where $\bar{T}_{\alpha\beta}$ is a geometrical tensor which must
be interpreted as the effective energy-momentum tensor for a
Riemannian geometry, but geometrically induced by the
non-metricity of the manifold
\begin{equation}
k \,\bar{T}_{\alpha\beta}= -K^{\mu}_{\mu\beta |\alpha}+K^{\mu}_{\alpha\beta |\mu}
-K^{\sigma}_{\mu\beta}K^{\mu}_{\sigma\alpha}+K^{\sigma}_{\alpha\beta}K^{\mu}_{\sigma\mu}
+\frac{1}{2}g^{\lambda\gamma}\left(K^{\mu}_{\mu\gamma |\lambda}-K^{\mu}_{\lambda\gamma |\mu}
+K^{\sigma}_{\mu\gamma}K^{\mu}_{\sigma\lambda}-K^{\sigma}_{\lambda\gamma}K^{\mu}_{\sigma\mu}\right)g_{\alpha\beta}.
\label{tn6}
\end{equation}
In absence of torsion, the effective contortion tensor
$K^{\sigma}_{\mu\nu}$ is due exclusively by the non-metricity
contribution:
\begin{equation}\label{tn7}
K^{\sigma}_{\mu\nu}=-\frac{1}{2}g^{\lambda\sigma}\left(N_{\mu\lambda\nu}+N_{\lambda\nu\mu}-N_{\mu\nu\lambda}\right).
\end{equation}
Physically we can interpret the equations (\ref{tn5}) as the
Einstein field equations with a cosmological constant term and a
geometrically induced tensor $\bar{T}_{\mu\nu}$, playing the role
of an energy-momentum tensor associated with matter sources. This
is a kind of induced matter theory where matter has a geometrical
origin, but in this case it is not induced by a foliation on an
extra-dimension, but on the non-metricity of the Weylian
connection. One problem with the non-metricity is related with the
integrability of the space-time. In simple words, non-metricity
leads to the atomic second clock effect. However, one geometry
with non-metricity free of this problem is the well known
Weyl-Integrable geometry. In the next section we will consider
this kind of space-time geometry.

\section{Weyl-Integrable manifold and gravitational waves from $\Lambda_1$}

We shall focus in this section on the particular case of a
Weyl-Integrable non-metricity \cite{07}. The non-metricity tensor
$N_{\sigma\mu\nu}$ and the non-metricity contortion tensor
$K^{\mu}_{\alpha\beta}$, are given by the expressions
\begin{eqnarray}\label{w1}
&& N_{\sigma\mu\nu}=g_{\mu\nu ;\sigma}=\varphi_{,\sigma}g_{\mu\nu},\\
\label{w2}
&& K^{\mu}_{\alpha\beta}=g_{\alpha\beta}\varphi^{,\mu}-\delta^{\mu}_{\alpha}\varphi_{,\beta}-\delta^{\mu}_{\beta}\varphi_{,\alpha},
\end{eqnarray}
where $\varphi (x)$ is a scalar field known in the literature as the Weyl scalar field.

The field equations (\ref{tn5}) are still valid for this
particular non-metricity, and the equation (\ref{tn21}) now reads
\begin{equation}\label{w3}
W^{\alpha}\varphi_{,\alpha}=\Lambda_1 g_{\alpha\beta}\delta g^{\alpha\beta},
\end{equation}
where $W^{\mu}$ is given by
\begin{eqnarray}
W^{\mu}&=&g^{\alpha\beta}\left[\frac{g^{\lambda\mu}}{2}\left\lbrace\delta g_{\alpha\lambda\,,\beta}
+\delta g_{\lambda\beta\,,\alpha}-\delta g_{\beta\alpha ,\lambda}\right\rbrace  -\frac{\delta g^{\lambda\mu}}{2}
\left[ g_{\alpha\lambda\,,\beta}+ g_{\lambda\beta\,,\alpha}- g_{\beta\alpha\,,\lambda}-g_{\alpha\lambda}\,
\varphi_{,\beta}- g_{\lambda\beta}\,\varphi_{,\alpha}\right.\right.\nonumber \\
&+& \left. g_{\beta\alpha}\,\varphi_{,\lambda}\right]-\frac{g^{\lambda\sigma}}{4}(\delta g_{\lambda\sigma\,,\beta}\,
\delta^{\mu}_{\alpha}+\delta g_{\lambda\sigma\,,\alpha}\,\delta^{\mu}_{\beta})
+ \left.\frac{\delta g^{\lambda\sigma}}{4}(g_{\lambda\sigma\,,\beta}\,\delta^{\mu}_{\alpha}+g_{\lambda\sigma\,,\alpha}\,\delta^{\mu}_{\beta}
-g_{\lambda\sigma}\,\varphi_{,\beta}\,\delta^{\mu}_{\alpha}-g_{\lambda\sigma}\,\varphi_{,\alpha}\,\delta^{\mu}_{\beta})\right].\nonumber\\
\label{w4}
\end{eqnarray}
Using the fact that $^{(W)}\Box g_{\alpha\beta}=g_{\alpha\beta ;\nu}\,^{;\nu}=g_{\alpha\beta}(2\varphi_{,\lambda}
\varphi_{,\rho}+\varphi_{,\rho ;\lambda})g^{\lambda\rho}$, and the expression (\ref{w4}), we obtain that the equation (\ref{w3}) can be written in the form
\begin{eqnarray}
&&\frac{1}{2}\Lambda_1 g^{\mu\nu}\left[\left(16\varphi_{,\alpha}\varphi^{,\alpha}-
4 g^{\lambda_1\rho}\left(2\varphi_{,\lambda_1}\varphi_{,\rho}+\varphi_{,\lambda_1 ;\rho}\right)
\right)\delta g_{\mu\nu}+\,^{(W)}\Box\delta g_{\mu\nu}\right]\nonumber \\
&& =\,^{(W)}\Box\Phi +4\left(2\varphi_{,\lambda_1}\varphi_{,\rho}+\varphi_{,\lambda_1 ;\rho}
\right)\left(W^{\lambda_1 ;\rho}-2W^{\lambda_1}\varphi^{,\rho}\right),\label{w5}
\end{eqnarray}
where we have introduced the scalar field $\Phi=W^{\mu}\varphi_{,\mu}$. It can be easily
seen that the condition $2\varphi_{,\mu}\varphi_{,\nu}+\varphi_{,\mu ;\nu}=0$ is valid when
the Weyl scalar field $\varphi$ satisfies the formula: $\bar{\Box}\varphi=0$. Thus, when the field $\varphi$ obeys a Riemannian wave equation,
the expression (\ref{w5}) reduces to
\begin{equation}\label{w6}
\frac{1}{2}\Lambda_1 g^{\mu\nu}\left[16\varphi_{,\alpha}\varphi^{,\alpha}\delta g_{\mu\nu}+\,^{(W)}\Box\delta
g_{\mu\nu}\right]=\,^{(W)}\Box\Phi,
\end{equation}
which must be separated to obtain particular
solutions
\begin{eqnarray}\label{wave31}
16 |v|^2\,\delta g_{\mu\nu} + \,^{(W)}\Box\,\delta g_{\mu\nu}=0,\\\nonumber ^{(W)}\Box\Phi=0,
\end{eqnarray}
with $|v|^2=\varphi_{,\alpha}\varphi^{,\alpha}$ for the Weyl
vector field in any point. With the help of the equations
(\ref{a4}) and (\ref{w2}), the expression (\ref{wave31}) can be
separated into a Riemannian part plus a part that depends of the
Weyl scalar field in the form
\begin{eqnarray}\label{wave32}
\bar{\Box}\,\delta g_{\mu\nu}=10 |v|^2\Phi,\\\nonumber \bar{\Box}\Phi=2\Phi_{,\nu}\varphi^{,\nu},
\end{eqnarray}
which must be viewed as an effective gravitational wave \cite{08}
with a massive term originated in the Weyl field. In the case in
which $\varphi=0$ ($|v|^2=0$), the problem reduces to a family of
homogeneous equations for Riemannian geometry.

\subsection{Massless Riemannian gravitons}

As was point out in last paragraph, we could obtain an homogeneous
Riemannian wave equation in the case with $\varphi=0$, but this is
not the only way. We see that (\ref{w6}) directly conduces to
\begin{equation}\label{wave4}
-10\,\Phi\,  |v|^2 + \frac{1}{2}\Lambda_1 g^{\mu\nu}\,\bar{\Box}\,\delta g_{\mu\nu}=\,\bar{\Box}\,\Phi-2 \Phi_{,\mu}\varphi^{,\mu}.
\end{equation}
If we set the particular solution with $\Phi=5\, e^{\varphi}$ and
$\Phi_{,\mu}\varphi^{,\mu}=5\,\Phi\,|v|^2$, hence the equation
(\ref{wave4}), must be rewritten as
\begin{eqnarray}\label{wave6} \bar{\Box}\,\delta g_{kc}=0,
\end{eqnarray}
plus the condition $\bar{\Box} \varphi=0$ which we already imposed
to reduce (\ref{w5}). Notice that the equation (\ref{wave4})
takes into account the simplest case corresponding to $\Phi_{,\, \nu}=0$, which
implies $|v|=0$. In this case this expression can be reduced to the particular form of
(\ref{wave6}), with the additional condition: $\bar{\Box} \Phi=0$.

\subsection{Massless Weyl-like graviton}

To study Weyl massless gravitons, we must take into account
(\ref{w6}), with the condition $2\, v_{\nu}\, v_{\mu}+v_{\nu;\mu}=0$,
in order to obtain solutions that make possible
$^{(W)}\Box\,\delta g_{ab}=0$, under the corresponding equation
for the Riemannian geometry:
\begin{equation}\label{wave7}
\frac{1}{2}\,g^{\mu\nu}\,^{(R)}\Box \delta g_{\mu\nu}=\{-10\,\Lambda_1\,g^{\mu\nu}|v|^2-2\,{g^{\mu\nu}}_{,\alpha}\,\varphi^{,\alpha}
-2\,g^{\alpha\beta}\,\varphi^{,\gamma}(\Gamma^{\nu}_{\beta\gamma}\delta^{\mu}_{\alpha}+\Gamma^{\nu}_{\alpha\gamma}\delta^{\mu}_{\beta})\}\,\delta g_{\mu\nu}.
\end{equation}
We see that
\begin{eqnarray}\label{z}
g^{\alpha\beta}\,\varphi^{,\gamma}(\Gamma^{\nu}_{\beta\gamma}\delta^{\mu}_{\alpha}
+\Gamma^{\mu}_{\alpha\gamma}\delta^{\nu}_{\beta})=-g^{\mu\nu}\,|v|^2-{g^{\mu\nu}}_{,\mu}\,\varphi^{,\mu},
\end{eqnarray}
which once evaluated in (\ref{wave7}), gives us a massive
Klein-Gordon equation on the Riemannian manifold
\begin{eqnarray}\label{wave8}
\bar{\Box}\,\delta g_{\mu\nu}+f(\Lambda_1)\,|v|^2\,\delta g_{\mu\nu}=0,
\end{eqnarray}
with
\begin{eqnarray}\label{z1}
f(\Lambda_1)=\frac{4-20\Lambda_1}{\Lambda_1+1}.
\end{eqnarray}
It is easy to see that for $\Lambda_1 \sim
\frac{1}{t^2}$, we obtain
\begin{eqnarray}\label{z2}
f(\Lambda_1)_{t \rightarrow \infty}\longrightarrow 4,
\end{eqnarray}
which in the asymptotic limit for (\ref{wave8})
\begin{eqnarray}\label{wave9}\bar{\Box}\,\delta g_{\mu\nu}+\,4\,|v|^2\,\delta g_{\mu\nu}=0,
\end{eqnarray}
corresponding to a massive Riemannian graviton with the mass term originated by the Weyl vector:
\begin{eqnarray}\label{z3}
m^2=4\,|v|^2=4\,\varphi_{,\alpha}\,\varphi^{,\alpha}.
\end{eqnarray}
It is easy to see that the equations (\ref{wave7}) and (\ref{w6}),
imply that
\begin{eqnarray}\label{wave10}
^{(W)}\Box\,\delta g_{\mu\nu}=0,
\end{eqnarray}
which is linked to a non massive Weyl graviton.

The present
formalism admits both kind of solutions: massive and massless
non-Riemannian gravitons. This versatility enables us to avoid the
problems of massive gravitons by dealing only with effective
Riemannian massless gravitons (which comes from a massive
non-Riemannian one), and make possible the study of "effective" massive Riemannian gravitons
which are really massless in a non-Riemannian sense. In both cases imply a new kind
of gravitational waves with source in non-metricity, which is a
difference with the usual GR gravitational waves.

\section{Gravitational waves from $\Lambda_2$}

In the previous section we have examined gravitational waves in
Weyl geometry using (\ref{tn21}), which is associated to the third
term of (\ref{a13}). Now we shall see that the second term of
(\ref{a13}), which also contributes to the total cosmological
constant, allows us to construct gravitational waves in a
different manner \cite{06}. Although we have supposed that the
surface term in (\ref{super}) is nonzero, we must notice that in
general, it is commonly considered in the literature as null.
However, this assertion do not implies that
$W^{\mu}_{;\mu}=g_{\mu\nu}\,W^{\mu;\nu}$ must be identically zero
in the inner of the manifold. Hence, one always can define a
geometrical field $\psi_{ab}$, according to
\begin{eqnarray}\label{psi}
\psi_{\alpha\beta;\gamma\delta}=\frac{1}{4}\,g_{\gamma\delta}\,^{(W)}W_{\alpha;\beta},
\end{eqnarray}
such that
\begin{eqnarray}\label{psi1}
^{(W)}\Box \psi_{\alpha\beta}=\,^{(W)}W_{\alpha;\beta}.
\end{eqnarray}
This is a wave equation with sources originated in the surface
term associated to $\Lambda_2$. Using the equations (\ref{a4}) and
(\ref{w1}), we obtain
\begin{equation}\label{www}
^{(W)}W^{\nu}=(\delta\{^{\nu}_{\mu\rho}\}-\delta\{^{\kappa}_{\kappa\rho}\}\delta^{\nu}_{\mu})g^{\mu\rho}
+(\delta K^{\nu}_{mr}-\delta K^{\kappa}_{\kappa\rho}\delta^{\nu}_{\mu})g^{mr}=\bar{W}^{\nu}+{\Delta W}^{\nu},
\end{equation}
with
$\bar{W}^{\nu}=(\delta\{^{\nu}_{\mu\rho}\}-\delta\{^{\kappa}_{\kappa\rho}\}\delta^{\nu}_{\mu})g^{\mu\rho}$
and  ${\Delta W}^{\nu}:=(\delta K^{\nu}_{\mu\rho}-\delta
K^{\kappa}_{\kappa\rho}\delta^{\nu}_{\mu})g^{\mu\rho}$. This last
is null when the generalized contortion tensor is null (i.e. on a
Riemannian manifold). Obviously
$^{(W)}W_{\alpha;\beta}=(g_{\nu\alpha}\,^{(W)}W^{\nu})_{;\beta}$.
On the other hand, we have
\begin{eqnarray}\label{www1}
^{(W)}W^{\nu}_{;\nu}=\,\bar{W}^{\nu}_{|\nu}+\,{\Delta W}^{\nu}_{|\nu}
+\,\bar{W}^{\nu}\,K^{\mu}_{\nu\mu}+\,{\Delta W}^{\nu}\,K^{\mu}_{\nu\mu},
\end{eqnarray}
with $\bar{W}^{\nu}_{|\nu}=\phi$ in the equation (10) of
\cite{06}, relating the divergence of the Weyl field
$^{(W)}W^{\nu}$ with the scalar field employed in such article. In
our case it is associated to the Weyl geometry, and the field
$^{(W)}W^{\nu}$ must be written in terms of
$\Phi=W^{\nu}\,v_{\nu}$:
\begin{eqnarray}\label{www2}^{(W)}W_{\alpha;\beta}=\,\bar{W}_{\alpha|\beta}+(\Phi_{,\beta}
+4\,\Phi\,\varphi_{,\beta})\frac{\varphi_{,\alpha}}{|v|^2},
\end{eqnarray}
after supposing the Weyl vector field is nonzero. Then,
(\ref{psi1}) becomes
\begin{eqnarray}\label{www3}
^{(W)}\Box \psi_{\alpha\beta}=\,\bar{W}_{\alpha|\beta}+(\Phi_{,\beta}+4\,\Phi\,\varphi_{,b})\frac{\varphi_{,\alpha}}{|v|^2},
\end{eqnarray}
which clearly reduces to
\begin{eqnarray}\label{www4}
\bar{\Box} \psi_{\alpha\beta}=\,\bar{W}_{\alpha|\beta},
\end{eqnarray}
in absence of non-metricity.

\section{Massless Gravitons from $\Lambda_1$ in pre-inflation}\label{eje} 

We consider the case of an expanding universe with cosmological constant $\Lambda_1 \neq 0$, such that the
metric tensor is $[g]_{\mu\nu} = {\bf
diag}[1, -a^2(t), -a^2(t), -a^2(t)]$. An interesting case is that of an pre-inflationary universe, with
a Hubble parameter
\begin{equation}\label{h}
H(t) = \sqrt{\frac{2\Lambda}{3}} \,
\tanh{\left[2\sqrt{\frac{2\Lambda}{3}}\, t\right]},
\end{equation}
such that the scale factor of the universe is
\begin{equation}\label{aa} a(t)=
\frac{a_0}{\left[1-\tanh^2{\left(2\sqrt{\frac{2\Lambda}{3}}\,
t\right)}\right]^{1/4}},
\end{equation}
with $a_0 =\sqrt{\frac{3}{2\Lambda}}$. This model describes an universe with a Hubble
parameter that increases from a null value to an asymptotically
constant value $\left.H(t)\right|_{t\gg G^{1/2}}
\rightarrow\sqrt{\frac{2}{3} \Lambda}$, describing the creation of
the universe and its transition from a static state to an
accelerated de Sitter inflationary expansion. This metric was
proposed and studied in a non-metricity free scenario in \cite{06}
and we follow his method to solve eq. (\ref{wave6}):
\begin{eqnarray}\nonumber \bar{\Box}\,\delta g_{kc}=0,
\end{eqnarray}
we propose an expansion of the form
\begin{eqnarray}\label{gww}
\delta g_{ab}(t,\vec{x}) &=& \frac{1}{(2\pi)^{3/2}}\,\sum_{M=+,\times} \,\int d^3 k \,\, e^M_{ab}(\hat{z})\nonumber \\
&\times & \left[ A_k \, e^{i\vec{k}.\hat{z}}  \, \chi_c(t) +
A^{\dagger}_k \, e^{-i\vec{k}.\hat{z}}  \, \chi^*_c(t)\right],
\end{eqnarray}
where index $M=+, \times$ denote the transverse polarizations
$+,\times$, on the plane normal to $\vec{k}$ and $e^M_{ab}$ are
the components of the polarization tensor, such that $e^M_{ab} \,
\bar{e}^{ab}_{M'} = \delta^M_{M'}$. We work in the frame in which
$\vec{k}$ is in the $\hat{z}$ direction, and the polarizations are
\begin{equation}
e^+_{ab}=\left(\begin{array}{ll} 1 & \,\,\,0 \\ 0 & -1
\end{array}\right)_{ab}, \qquad \qquad
e^{\times}_{ab}=\left(\begin{array}{ll} 0 & 1 \\ 1 & 0
\end{array}\right)_{ab},
\end{equation}
with $a,b$ spanning the $(x,y)$ plane. We shall use the TT gauge
in order to solve the gravitational waves equations, such gauge is
represented by next conditions
\begin{equation}
\delta g_{0\mu}=0, \qquad \delta g^{i}_{\,\, i} =0, \qquad
\bar{\nabla}^j \delta g_{ij}=0.
\end{equation}
The equation of motion for the modes $\chi_k(t)$ is
\begin{equation}\label{chi}
\ddot\chi_c(t) + 3 \frac{\dot a}{a} \dot\chi_c(t) +
\frac{k^2}{a(t)^2} \chi_c(t)=0.
\end{equation}
The annihilation and creation operators $A_{k}$ and
$A_{k}^{\dagger}$ satisfy the usual commutation algebra
\begin{equation}\label{m5}
\left[A_{k},A_{k'}^{\dagger}\right]=\delta
^{(3)}(\vec{k}-\vec{k'}),\quad
\left[A_{k},A_{k'}\right]=\left[A_{k}^{\dagger},A_{k'}^{\dagger}\right]=0.
\end{equation}
We obtain the normalization condition for the modes
$\chi_{c}(\tau)$:
\begin{equation}\label{m6}
\chi_{c}(t) \frac{d{\chi}^*_{c}(\tau)}{d\tau} - \chi^*_{c}(t)
\frac{{\chi}_{c}(\tau)}{d\tau} = i
\,\left(\frac{a_0}{a(\tau)}\right)^{3},
\end{equation}
with $\tau =b\,t$, where
$b=\sqrt{\frac{2\Lambda}{3}}=\frac{1}{a_0}$. The asterisk in
(\ref{m6}) denotes the complex conjugated. For the case in which
the Hubble parameter and the scale factor are given respectively
by (\ref{h}) and (\ref{aa}), the general solution for the
amplitudes $\chi_c(\tau)$ is
\begin{eqnarray}
\chi_c(\tau) &=& C_1\, \frac{\sinh{(\tau)}}{\sqrt{2\cosh^2{(\tau)}-1}} \nonumber \\
 &\times &{\rm Hn}\left[-1,\frac{c^2-1}{4};0,\frac{1}{2},\frac{3}{2},\frac{1}{2};-\tanh^2{(\tau)}\right] \nonumber \\
&+& C_2\, \frac{\cosh{(\tau)}}{\sqrt{2\cosh^2{(\tau)}-1}} \nonumber \\
&\times & {\rm Hn}\left[-1,\frac{c^2+1}{4};-\frac{1}{2},0,\frac{1}{2},\frac{1}{2};-\tanh^2{(\tau)}\right] , \nonumber \\
\end{eqnarray}

where ${\rm
Hn}[a,q;\alpha,\beta,\gamma,\delta;z]=\sum^{\infty}_{j=0} c_j\,
z^j$ is the Heun function. We can do a series expansion in both
sides of (\ref{m6}) according to
\begin{equation}
\chi_{c}(t) \frac{d{\chi}^*_{c}(\tau)}{d\tau} - \chi^*_{c}(t)
\frac{{\chi}_{c}(\tau)}{d\tau} -
i\,\left(\frac{a_0}{a(\tau)}\right)^{3} = \sum^{\infty}_{N=1}
f_N(c)\, \tau^N=0,
\end{equation}
where $f_N(c^{(N)}_n)=0$, for each $N$. There are $2N$ modes for
each $N$-th order of the expansion, we obtain conditions for
coefficients $C_1$ and $C_2$, and the  values for wave-number $k$.
From the zeroth order expansion (in $\tau$), we obtain that
$C_2=i\,C_1/2$. Hence, we shall choose $C_1=1$ and $C_2=i/2$.

The serie is infinite so that there are infinite values of
quasi-normal modes with projection on the plane orthogonal to the
direction of propagation with zero norm according to $\hat{z}$:
$\vec{k} =c^{(N)}_n \hat{e^1} + (\pm i c^{(N)}_n) \hat{e^2} + k
\hat{e^3}$, with complex values $c^{(N)}_n$ which provide us the
quantization of tensor fields $\delta g_{ab}$. Notice that
$\|\vec{k}\|^2=k^2$, so that the norm of the polarization vectors
on the plane ($\hat{e^1},\hat{e^2}$), is zero. When the
polarization modes which are purely imaginary or purely reals
correspond to values with polarization $+$. On the other hand,
modes with complex-$c^{(N)}_n$ correspond to those with
polarization $\times$. In the table we have included the square
wave-numbers for the first five orders of the expansion:
\begin{small}
\begin{center}
\begin{tabular}{| r | l  r | l | r | l | r |}
\hline \small Square wavenumbers $c^{(N)}_n$  & \small
$\left(c^{(N)}_n\right)^2 $ values    \\ \hline \small
$\left(c^{(1)}\right)^2$ & \small -0.5   \\  \hline
$\left(c^{(2)}\right)^2$ & \small 6.57  & \small -0.57   \\ \hline
\small $\left(c^{(3)}\right)^2$ & \small -53.78  & \small 0.01 +
1.25 i & \small 0.01 - 1.25 i \\ \hline \small
$\left(c^{(4)}\right)^2$ & \small 9.60 & \small -1.29 & \small
9.85 + 9.40 i  & \small 9.85 - 9.40 i  \\ \hline \small
$\left(c^{(5)}\right)^2$ & \small 5.93 & \small 12.72 & \small
-1.34 & \small 13.84 + 18.30 i & \small 13.84 - 18.30 i  \\ \hline
\end{tabular}\label{tt}
\end{center}
\end{small}
\vskip .2cm
which is provided in \cite{06} and formally solves eq.
(\ref{wave6}). We must remark that although eq. (\ref{wave6}) is
formally equal to eq. (42) in \cite{06} the origin of our equation
is slightly different since we got eq. (\ref{wave6}) as the
expression for an effective gravitational wave from a Weyl
space-time which directly describes the tensorial perturbation of
a background FRW metric.

\section{Final Comments}

We have studied a new formalism to describe gravitational wave dynamics on a Weylian
manifold. The case in which the geometry is free of torsion has been examined.
This case will be
considered in a further work. Of course, due to the fact the
nature of the manifold is Weylian, the non-metricity and the
boundary terms must be taken into account. The fist case was
studied in Sect. III and provide us an integrable treatment of the
Weylian manifold. The second case was studied in Sect. IV and is
important because nontrivial boundary terms could be responsible
for the cosmological constant produced by a source inside a 3D
Gaussian-like hypersurface that encloses that source. This source
also could be responsible for the production of gravitational
waves. To finalize, it is important to notice that in the paper
\cite{007}, the authors obtain a gauge-invariant relativistic
quantum geometry by using a Weylian-like manifold with a geometric
scalar field, which provides a gauge-invariant relativistic
quantum theory, in which the quantum algebra of the Weylian-like
field depends of the observers. Is important to notice that, in
the context of \cite{007}, the fields $^{(W)}{W}_{\alpha}$ and
$\bar{W}_{\alpha}$ must be interpreted as variations, the first
one linked to the quantum geometry described by the Weylian-like
manifold, and the second one related to classical process related
to the Riemannian manifold, which must be zero. In order to make
our formulation compatible with the ideas of such work, we must
impose the condition $\bar{W}_{\alpha}=0$, which conduces from
(\ref{www2}) and (\ref{www3}), to
\begin{eqnarray}\label{www33}
^{(W)}W_{\alpha;\beta}=(\Phi_{,\beta}
+4\,\Phi\,\varphi_{,\beta})\frac{\varphi_{,\alpha}}{|v|^2}=\,^{(W)}\Box
\psi_{\alpha\beta},
\end{eqnarray}
where the last equation represents an effect originated in the
$\Lambda_2$ frontier contribution and enterally due to
non-metricity, which is of quantum nature.

In Sct. (\ref{eje}), we have solved the particular example
of the gravitational wave associated to Riemannian massless 
gravitons over a FRW background describing a pre-inflationary
scenario. The normalization conditions for the modes impose that the wavenumbers 
for $\delta g_{\alpha\beta}$ are an infinity number of discrete complex values for the polarization modes, which can be real and imaginary 
for the $+$-modes and complex for the $\times$-modes.

\section*{Acknowledgements}

\noindent J. M. Romero and M. Bellini acknowledge UNMdP and
CONICET for financial support. J.E.M.A acknowledges CONACYT
M\'exico, Centro Universitario de Ciencias Exactas e Ingenierias
and Centro Universitario de los Valles, of Universidad de
Guadalajara for financial support.

\end{document}